\documentclass[german,english,12pt,twoside]{article}
\usepackage{babel}
\usepackage[latin2]{inputenc}
\usepackage{graphicx}

\input CoKon.sty
\begin{document}

\CoKonChapBeg
 
\CoKonChapterTitle{Spectroscopic studies of star forming regions}  

\CoKonChapterAuthor{M\'aria Kun}

\CoKonChapterAuthorAffil{Konkoly Observatory of the Hungarian Academy of Sciences\\
P.O. Box 67, H-1525 Budapest, Hungary}

\CoKonAbstract{ 
This paper reviews the results of studies of star forming regions, 
carried out at the Konkoly Observatory in the last two decades. The studies 
involved distance determination of star-forming dark clouds, search for 
candidate pre-main sequence stars, and determination of the masses and 
ages of the candidates by spectroscopic follow-up observations. 
The results expanded  the list of the well-studied 
star forming regions in our galactic environment. Data obtained by 
this manner may be useful in addressing
several open questions related to galactic star forming processes.}

\CoKonSection{Introduction}
\label{sec1}

The first instrument of the Piszk\'estet\H{o} Observatory, the 60/90/180\,cm
Schmidt telescope with two objective prisms has been a precious part of Detre's legacy. We inherited with this instrument a challenge to probe new paths of research, inspired by the attributes of the Schmidt telescope, and use techniques apparently very different from those of variable star researches, our most important scientific heritage.

The telescope with its field of view of 5~degrees was suitable for performing
homogeneous, moderate-accuracy photographic photometry 
and low-dispersion slitless spectroscopy of large number of stars 
simultaneously. Traditional topics for Schmidt telescopes equipped 
with objective prisms included studying the space distribution of 
the stars and the light-absorbing diffuse matter in our galactic neighbourhood. 
Detre's early paper (Dunst, 1929) testifies that this subject was 
close to his scientific interest at the beginning of his career.
Searching for objects of peculiar spectra in objective prism plates is 
another suitable project for this instrument. This kind of study has 
to be followed by slit spectroscopy of the selected objects, in order to
establish their real nature. Galactic star forming regions, extending over large volumes of space and containing large amount of interstellar gas and dust, as well as young stellar objects of emission spectra are suitable targets for both kinds of Schmidt surveys. During the past two decades several star forming regions were discovered and studied using the Schmidt telescope with the 5-degree objective prism.  The studies included distance 
determination and identifying optically visible pre-main sequence stars. 
The follow-up spectroscopic studies of the candidate pre-main sequence 
stars have been performed with larger telescopes available thanks to the 
OPTICON project. In this paper I introduce some star forming regions whose basic properties have been determined using the Schmidt telescope at Piszk\'estet\H{o}.

\CoKonSection{Star forming regions} 

\CoKonSubsection{Open issues of star formation}

Several important properties of the star forming processes can be understood 
only by comprehensive, large-scale, multi-wavelength studies of star forming regions. For instance, complete mapping and detailed photometric and spectroscopic studies of the young stellar population are required for establishing the stellar initial mass function, the age distribution of young stars, and studying the time scale of star formation. Both the star forming cloud and the stars born in it have to be mapped in order to study the 
efficiency of the star forming process and its propagation within the cloud.

Another group of open questions cannot be answered by detailed studies of a
single nearby star forming region, but requires comparison of several regions. Which properties of molecular clouds influence the time scale 
and efficiency of star formation, the clustered and isolated modes, and the
mass spectrum of the newborn stars? How does the early evolution of the stars depend on the environment of the star formation? Is star formation a fast and 
dynamic process governed by supersonic turbulence or rather a slow, 
quasistatic process controlled by static magnetic fields? 
To answer these questions, the stellar content of several star forming
regions has to be assessed, and then the mass and age distributions of young 
stellar objects over the whole area of the star forming region, as well as 
the spectral energy distribution, characterizing the 
circumstellar environments of the young stars have to be determined.

\CoKonSubsection{Scope of our researches}

In the 1980's the development of the millimetre-wave radio astronomy led to
the discovery of a large population of molecular clouds outside the galactic plane ($|b| > 10^\mathrm{o}$). Part of them are smaller and more transparent than their galactic plane siblings. The discovery of the new molecular clouds aroused new problems: How far are the newly discovered molecular clouds from us? Do they form low-mass stars? Objective prism observations of the molecular cloud regions and follow-up spectroscopy provide the most suitable means of finding the answers to the questions.

We selected for studies molecular clouds which have already been 
mapped in $^{12}$CO or/and $^{13}$CO.  Some of the target clouds also have shown evidence of low-mass star formation, e.g. embedded {\it IRAS\/} sources are associated with them. These clouds certainly contain a less conspicuous population of young pre-main sequence stars. The goal of our studies is to derive some basic properties of the selected star forming regions: to determine their distances,
to assess the number of pre-main sequence stars, as well as to estimate
their  mass and  age distributions. The first steps of the studies 
were based on objective prism observations using the 60/90/180\,cm 
Schmidt telescope of the Konkoly Observatory. Follow-up observations of
the candidate pre-main-sequence stars have been carried out with various optical telescopes.

\CoKonSection{Methods}

\vskip -3mm

\CoKonSubsection{Distance determination}

In determining distances to nearby molecular clouds their interactions
with the light of embedded or background stars can be utilized. While 
distances of normal stars can be determined from their apparent and absolute magnitudes (spectral types and colours), no properties of dark clouds
indicate how far they are located from us. This is especially true for nearby 
clouds, whose galactic orbital  velocities are close to the solar value.
Neither the young, embedded stars are good distance indicators, because
their luminosities strongly decrease during the pre-main sequence evolution.
Extinction of starlight by the dust content of the cloud, interstellar absorption lines in the spectra of background stars, and  embedded stars  
illuminating reflection nebulae can be used for distance determination. 
Distance is a basic data, which is important for determining the  size and mass of the cloud, and the absolute luminosities of young stars.

In order to determine the distances  of the selected clouds we examined
the cumulative distribution of the field star distance moduli 
along the line of sight to the clouds. If  $y=V-M_{\rm V}$ is the 
distance modulus of the stars, and $N(y)$ denotes the number of stars 
whose distance modulus is smaller than $y$ (i.e. brighter than
$V = M_{\rm V} + y$), then the distance modulus for distances smaller 
than that of the obscuring cloud can be written as 
$y=V-M_V=5\log\,r-5$, whereas behind the cloud it is $y=5\log\,r-5+A_V$, 
where $A_V$ is the visual extinction caused by the cloud.
Thus the presence of a cloud along the line of sight shows up as a distortion in the shape of the $logN(y)$ {\it vs.\/} $y$ curve ({\sl Wolf diagram}),
and its distance modulus can directly be read from the diagram. 

The absolute magnitudes of stars were estimated from their objective 
prism spectral types. The spectra were obtained with the Schmidt 
telescope of the Konkoly Observatory equipped with a UV-transmitting objective prism having a refracting angle of 5$^{\mathrm o}$ and a dispersion of 580\,\AA/mm at H$\gamma$. The field of view of the telescope was 19.5 square degrees. Absolute magnitudes of stars belonging to different spectral classes were taken from Allen (1973) and Cox (1999). For calculating the  apparent distance moduli $V-M_\mathrm{V}$ of the stars  the $V$ magnitudes listed in 
the {\sl Guide Star Catalog\/} were used.

In this manner the distances of the dark clouds L\,694 (Kawamura et~al., 2001), L\,1228, L\,1235, L\,1241, L\,1251, L\,1261 (Kun, 1998), L\,1333 (Obayashi et~al., 1998), and L\,1340 (Kun et~al., 1994) were determined. During this work spectral types of some 5000 stars have
been determined visually, on objective prism plates.

\begin{figure}[!h]
\centerline{
\includegraphics[width=8.cm]{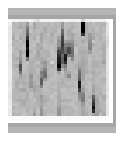}
\includegraphics[width=6.6cm]{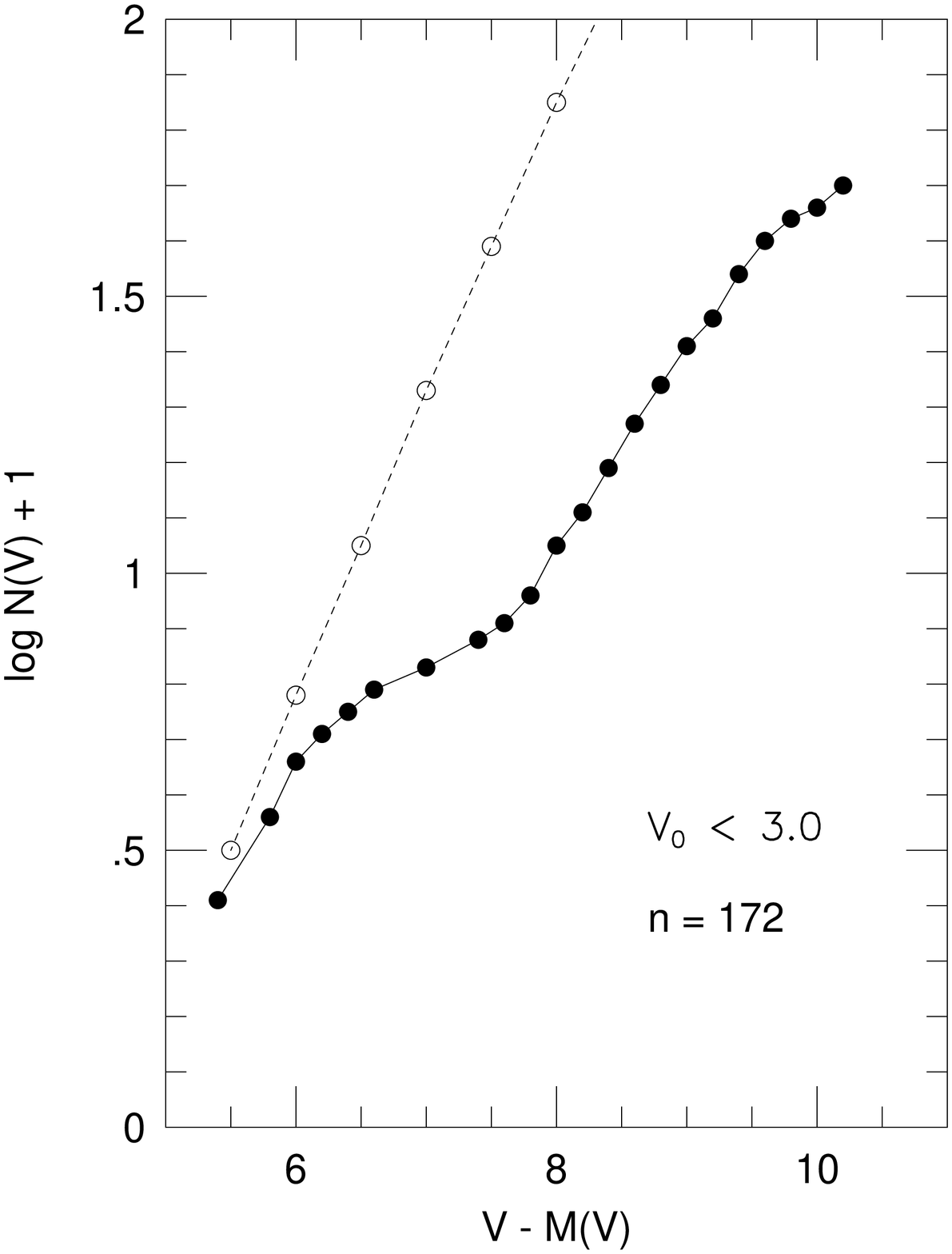}}
\caption{{\it Left}: Part of a blue-sensitive (emulsion type IIaO) objective prism plate obtained with the Schmidt telescope through the 5-degree prism. 
{\it Right}: Plot of $log N(V)$ vs. $V-M_V$ for the stars with $M_V < 3.0$. {\it N(V)\/} 
is the number of stars with apparent magnitudes brighter than {\it V\/} within 1 deg$^2$.
 The dashed line indicates the absorption-free reference curve. \label{fig1}}
\end{figure}

An example for the Wolf diagram, showing the distribution of stellar
distance moduli in the region of L1333, is shown in {\it Fig.~\ref{fig1}}. The number of stars is normalized to one square degree. Filled circles connected by solid line show the cumulative distribution of distance moduli determined for a field of 19 square degrees centered on the cloud, and the dashed line is the reference curve displaying the same distribution without extinction at the galactic latitude of the cloud, $+15^{\mathrm o}$. The error of this kind of distance determination can be estimated as $\pm$\,10 percent, from the accuracy of spectral classification and the GSC magnitudes, as well as from the number of stars involved in the curves (Obayashi~et~al.,~1998).

\CoKonSubsection{Search for candidate pre-main sequence stars}

I have been searching for pre-main sequence star candidates using objective
prism spectra taken with the Schmidt telescope through a red filter. The dispersion of the objective prism is about 2000\,\AA/mm at H$\alpha$. 
{\it Fig.~2} shows an example of a plate taken through a red (Schott~RG\,1) filter. {\it Fig.~\ref{fig_sp_prism}} shows examples of the objective prism spectra of various H$\alpha$ emission objects obtained with the Photometrics CCD camera installed on the Schmidt
telescope. The wavelength scale of the very low dispersion spectra
was established in possession of the geometric and optical properties of
the prism and camera, and using the atmospheric A-band at 7600\,\AA\ as 
reference wavelength. The quality of such a low dispersion slitless spectrum 
highly depends on the atmospheric conditions. This method is suitable for finding classical T~Tauri stars, displaying strong H$\alpha$ emission 
({\it EW\/}(H$\alpha$) $\ge $10\,\AA). 
In addition to the objective prism spectra, infrared point sources of 
the {\it IRAS\/} Point Source Catalog and Faint Source Catalog were used 
to find further pre-main sequence stars born in the selected clouds.

\begin{figure}
\resizebox{7cm}{!}{\includegraphics{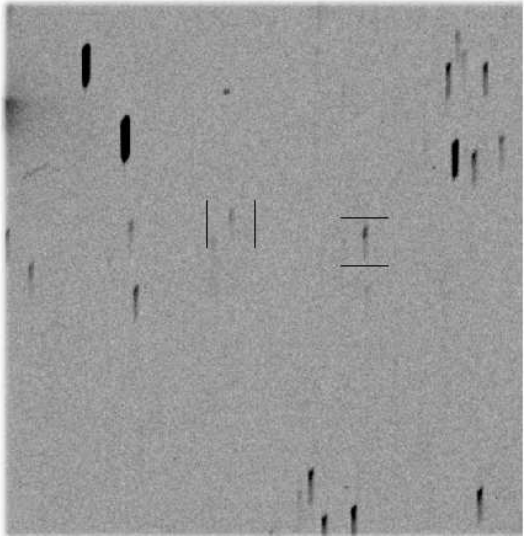}}
\vskip -56mm
\hfill
\parbox[b]{60mm}{
\caption{Objective prism image of the northern core of the dark cloud
Lynds~1251. Part of a plate taken on Kodak 098-02 
emulsion and through an RG1 filter with an exposure time 60~min. The 
images of three H$\alpha$ emission stars are framed.}}
\vskip 14mm
\label{fig_ha_prism}
\end{figure}

\begin{figure}[ht]
\centerline{\includegraphics[width=14cm]{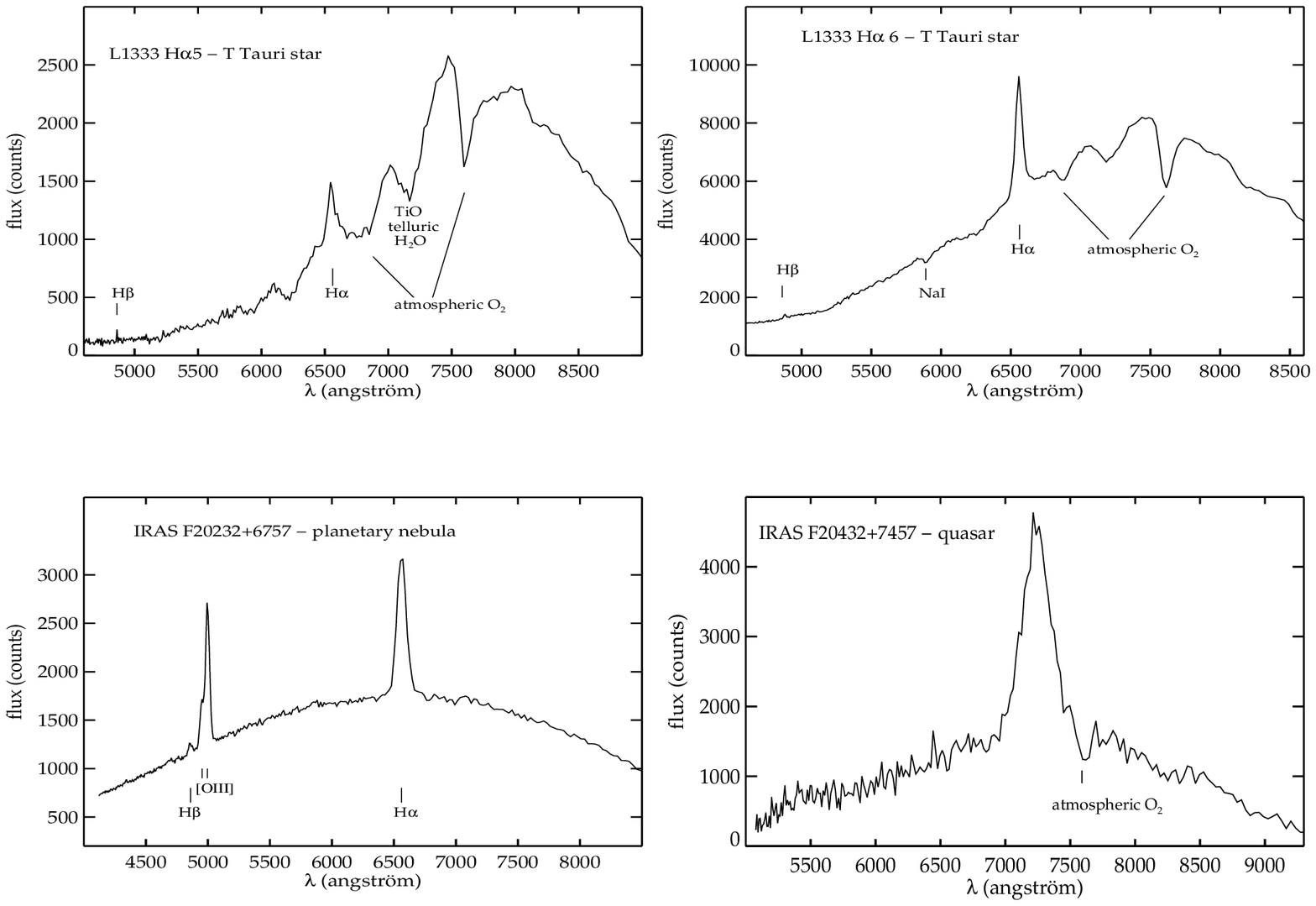}}
\vskip -35mm
\caption{Objective prism spectra of various objects shown up as H$\alpha$ 
emission stars. Position of the H$\alpha$ line and the {\it A\/}-band
of the atmospheric oxygen are indicated.}
\label{fig_sp_prism}
\end{figure}

\CoKonSubsection{Spectroscopic follow-up observations}
\label{sec3.3}

The real nature of the selected stars has to be established by medium-resolution spectroscopic follow-up observations. We performed spectroscopic observations of a few regions studied earlier with the objective prism using different telescopes and instruments: the {\it Intermediate Dispersion Spectrograph\/} on Isaac Newton Telescope, La Palma, {\it CAFOS\/} on the 2.2\,m telescope of Calar Alto Observatory, as well as {\it ALFOSC\/} on the Nordic Optical Telescope. The spectra were used for determining the spectral types of the stars and for establishing their pre-main sequence nature. Criteria for classification are described in Kirkpatrick et~al. (1991), Preibisch et~al. (2001), and Mart\'{\i}n \& Kun (1996). Several candidate pre-main sequence stars proved to be field stars without H$\alpha$ emission.
Nevertheless, a number of classical T~Tauri stars were found in molecular clouds where no pre-main sequence stars had been identified earlier. 

\CoKonSection{Results}

\CoKonSubsection{Low-mass star formation in small 
molecular clouds}
\label{sec4.1}

Star forming molecular clouds tend to be parts of giant molecular complexes. The small translucent clouds found at high galactic latitudes usually are not associated with prominent signposts of star formation, such as far-infrared point sources. Our objective prism search for pre-main sequence stars at high galactic latitudes in most cases had negative results: spectroscopic follow-up observations have not confirmed the pre-main sequence nature of the candidate stars picked up from the photographic plates (Mart\'{\i}n \& Kun, 1996). The following sections show a few exceptions.

\CoKonSubsubsection{The Lynds 134 complex} 
\label{sec4.1.1}

 The members of this complex are the small dark clouds  L\,134, L\,169, L\,183, L\,1780, located at a distance of 110\,pc from us (Franco, 1989). No star formation was found earlier in the L\,134 complex. The first star formation signpost in this region was the very low mass, young T~Tauri star (spectral type: M\,5.5IV) found during our survey (Mart\'{\i}n \& Kun, 1996) at an angular distance of 5~arcmin (0.16\,pc) from the dense core L\,183{\it i\/} (Laureijs et~al., 1995). The radial velocities of the cloud and the star are close to each other, suggesting that the star was born in the 
cloud a few million years ago. The pre-protostellar condensations 
revealed recently in the same cloud by far-infrared and submm observations 
(Lehtinen et~al., 2003) indicate ongoing star formation in L\,183.

We found a binary system of very low mass, weak-line T~Tauri stars (spectral types: M4V and M5IV) in the same region, to the west from the cometary cloud L\,1780. The morphology of the region suggests that both the cometary shape of L\,1780 and the formation of the low-mass stars were triggered by stellar winds form the Sco-Cen association (see T\'oth et~al., 2003).

\CoKonSubsubsection{IC\,2118}
\label{sec4.1.2}

\begin{figure*}[t]
\centerline{\includegraphics[width=7cm]{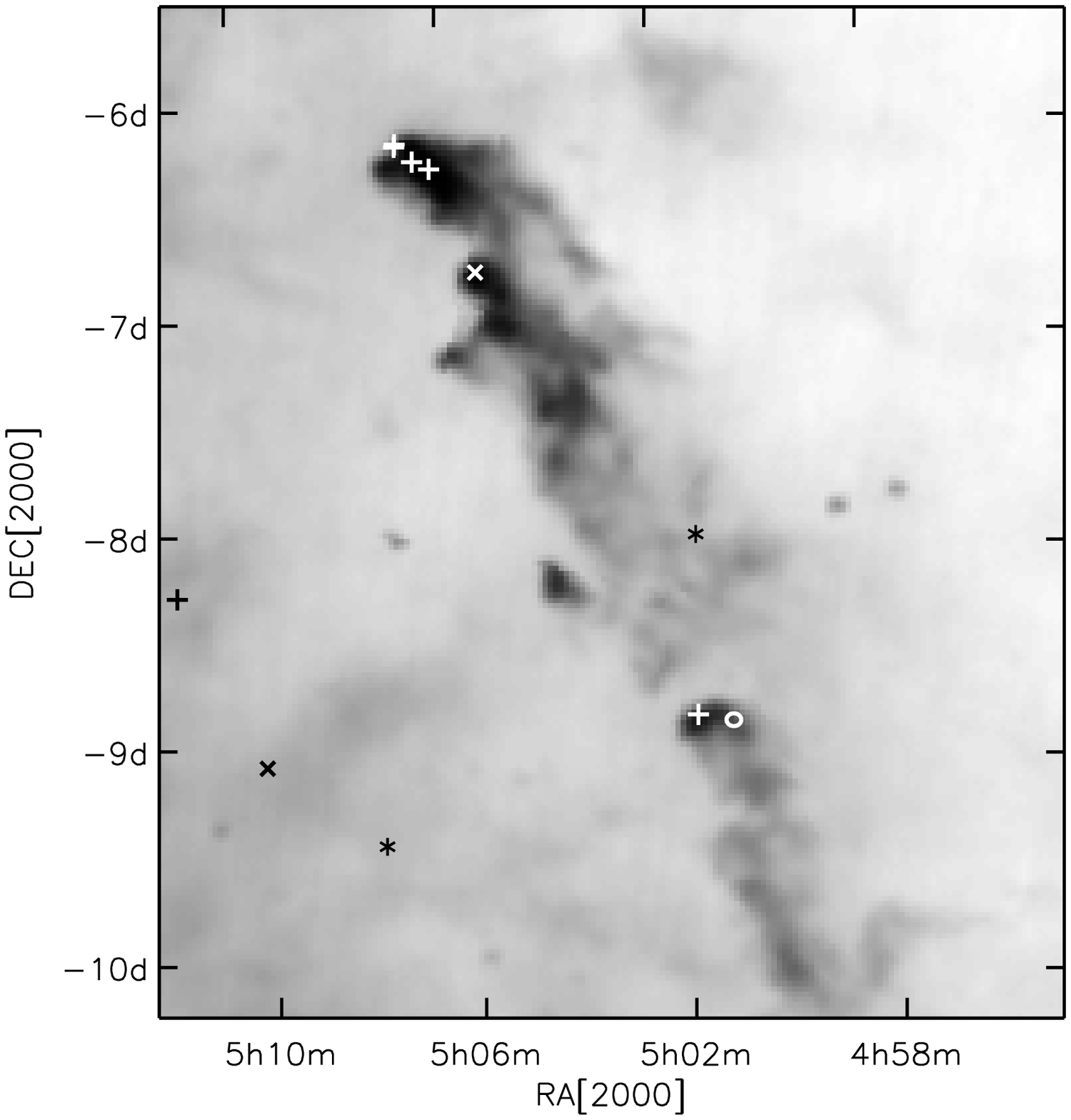}\includegraphics[width=9cm]{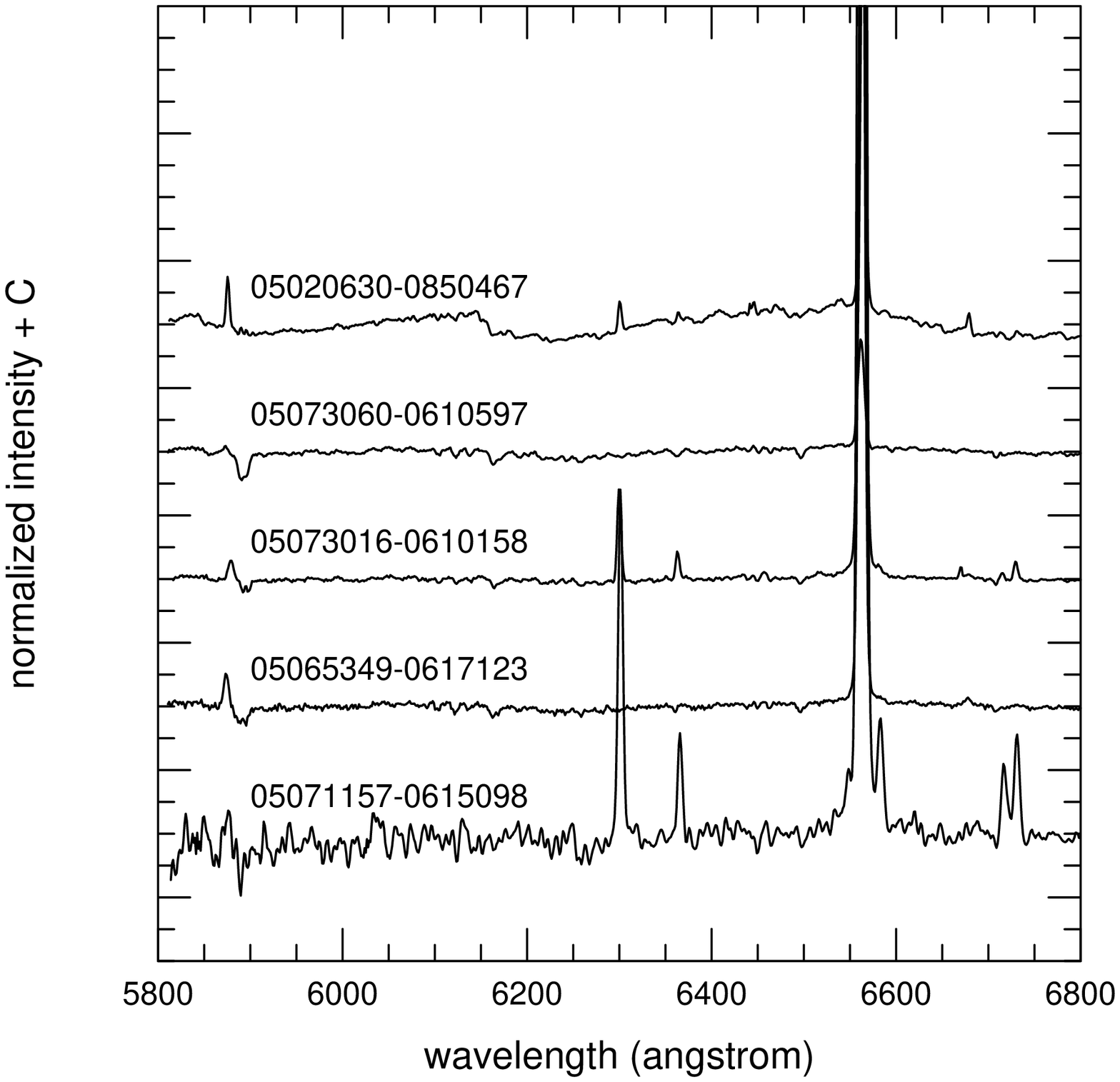}}
\vskip -5mm
\caption{{\it Left\/}: Distribution of classical T~Tauri stars on the 100$\mu$m IRAS map of IC\,2118.
{\it Right\/}: Optical spectra of classical T~Tauri stars in IC\,2118 over the wavelength 
region 5800--6800\,\AA.} \label{fig_icmap}

\end{figure*}

IC\,2118 is an extended reflection nebulosity at high galactic latitude 
($27^\mathrm{o} < b < 33^\mathrm{o}$), illuminated by $\beta$ Orionis. It 
is situated at a distance of 210\,pc from the Sun, and at some ten degrees to 
the west from the Orion star forming region. The bright nebula is associated by several small molecular clouds (Kun et~al., 2001), among others MBM~21 and 22 (Magnani et~al., 1995). Two young stellar object (YSO)-like {\it IRAS\/} sources, {\it IRAS\/}~04591$-$0856 and {\it IRAS\/} 05050$-$0614 are indicative of ongoing low-mass star formation in this region. Follow-up spectroscopic studies of our H$\alpha$ emission stars selected as pre-main sequence star candidates, performed with the {\it ALFOSC\/} spectrograph on Nordic Optical Telescope, resulted in the discovery of five classical T~Tauri stars (cTTS) projected on the molecular clouds (Kun et~al., 2004). Their distribution is shown in the left panel of {\it Fig.~\ref{fig_icmap}}, overlaid on the {\it IRAS\/} 100$\mu$m image of the region. The spectra and the JHK photometric data of the 2MASS All Sky Survey (Cutri et~al., 2003) made it possible to determine the effective temperature $T_\mathrm{eff}$ and luminosities $L/L_{\odot}$ of the young stars. The spectra of the T~Tauri stars of IC\,2118 are shown in the right panel of {\it Fig.~\ref{fig_icmap}}, and their positions in Hertzsprung--Russell diagram can be seen in {\it Fig.~5}.

\begin{figure}[!]
\vskip -1cm
\resizebox{10cm}{!}{\includegraphics{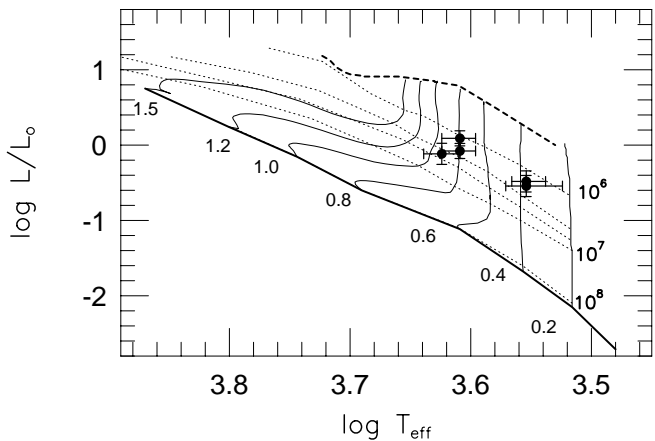}}
\vskip -6cm
\hfill
\parbox[b]{50mm}{
\caption{Location of the IC\,2118 pre-main-sequence stars in the 
Hertzsprung--Russell diagram, together with the pre-main-sequence 
evolutionary tracks and isochrones published by (Palla \& Stahler, 1999).} \label{fig_ichrd}}
\vskip 10mm
\end{figure}

The  molecular clouds associated with IC\,2118 are among the smallest 
known star forming clouds. They probably lie near the surface of the
{\it Orion--Eridanus Bubble\/}, being blown by the
stellar winds and supernova explosions of the massive stars
of Orion~OB1 during the past ten million years. 
Given the large line-of-sight distance between the clouds and
Orion~OB1, star formation in this region propagates not only from
the east to the west, but also towards us. The ages of
the pre-main sequence stars found in the clouds are compatible
with the assumption that star formation has been triggered by
the superbubble. The complicated geometry and wind history of the
OB association (Brown et~al., 1995) hinders any detailed
speculation on the exact position and age of the sources of
trigger. Recently observations by the Spitzer Space Telescope revealed 
several very low mass young stars in these clouds.

\vfill\eject
 
\CoKonSubsection{L\,1333}
\label{sec4.1.3}

L\,1333 is a small dark cloud in Cassiopeia. The basic properties of the cloud and its environment were studied by Obayashi et~al. (1998) within the framework of the collaboration between the Konkoly Observatory and the Radio Astronomy Laboratory of Nagoya University. We obtained a distance of 180\,$\pm$\,30\,pc for the cloud. $^{13}$CO and C$^{18}$O observations revealed that the dark cloud is a part of a filamentary molecular complex, consisting of small dense clumps separated by some 6~pc from each other along a narrow line. The total mass of the complex, determined from the molecular observations, is about 720\,M$_\odot$. Star formation is indicated by the protostellar-like {\it IRAS\/} source {\it IRAS\/}~02086+7600. Our objective prism and subsequent spectroscopic studies of L\,1333 (Kun et~al., 2006) revealed four classical T~Tauri stars in the region of L\,1333, associated with the {\it IRAS\/} sources F02084+7605, 02103+7621, and 02368+7453. One of the {\it IRAS\/} sources, {\it IRAS\/} 02103+7621 (OKS\,H$\alpha\,6$ in Obayashi et~al., 1998) proved to be a visual binary, whose both components are cTTS, separated by about 1.8~arcsec (corresponding to $\sim$ 320\,AU at 180\,pc).

\begin{figure*}[ht]
\centerline{\includegraphics[width=14cm]{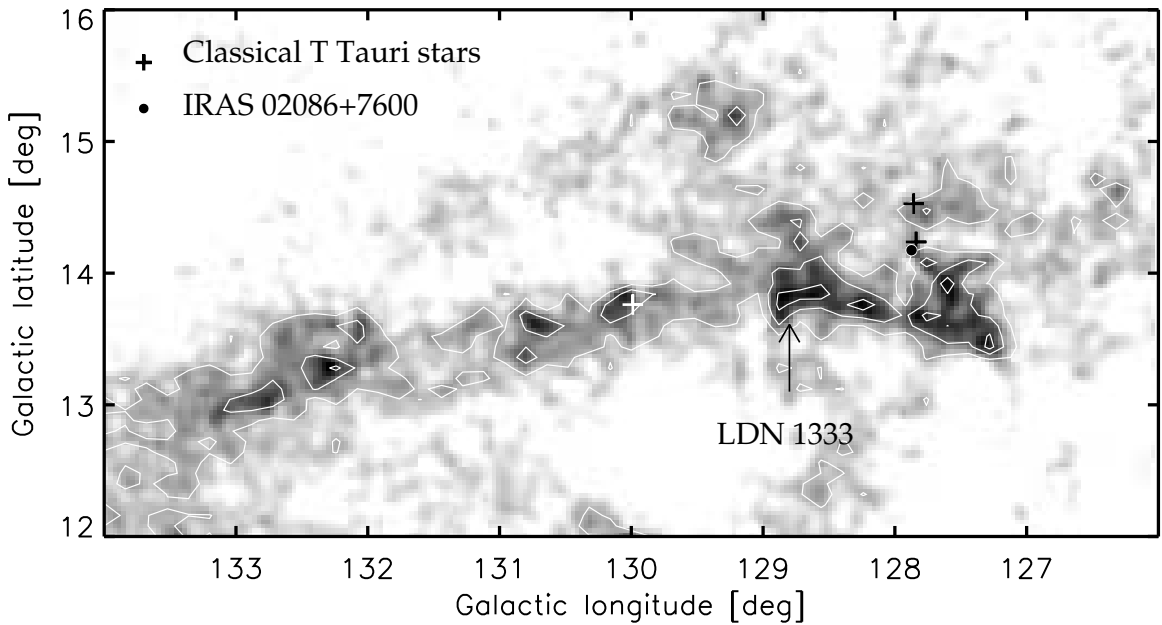}}
\vskip -5mm
\caption{Distribution of the visual extinction  and
the young stars in the region of L\,1333.}
\label{fig_map1333}
\end{figure*}

\begin{figure*}[!]
\centering{
\includegraphics[width=7cm]{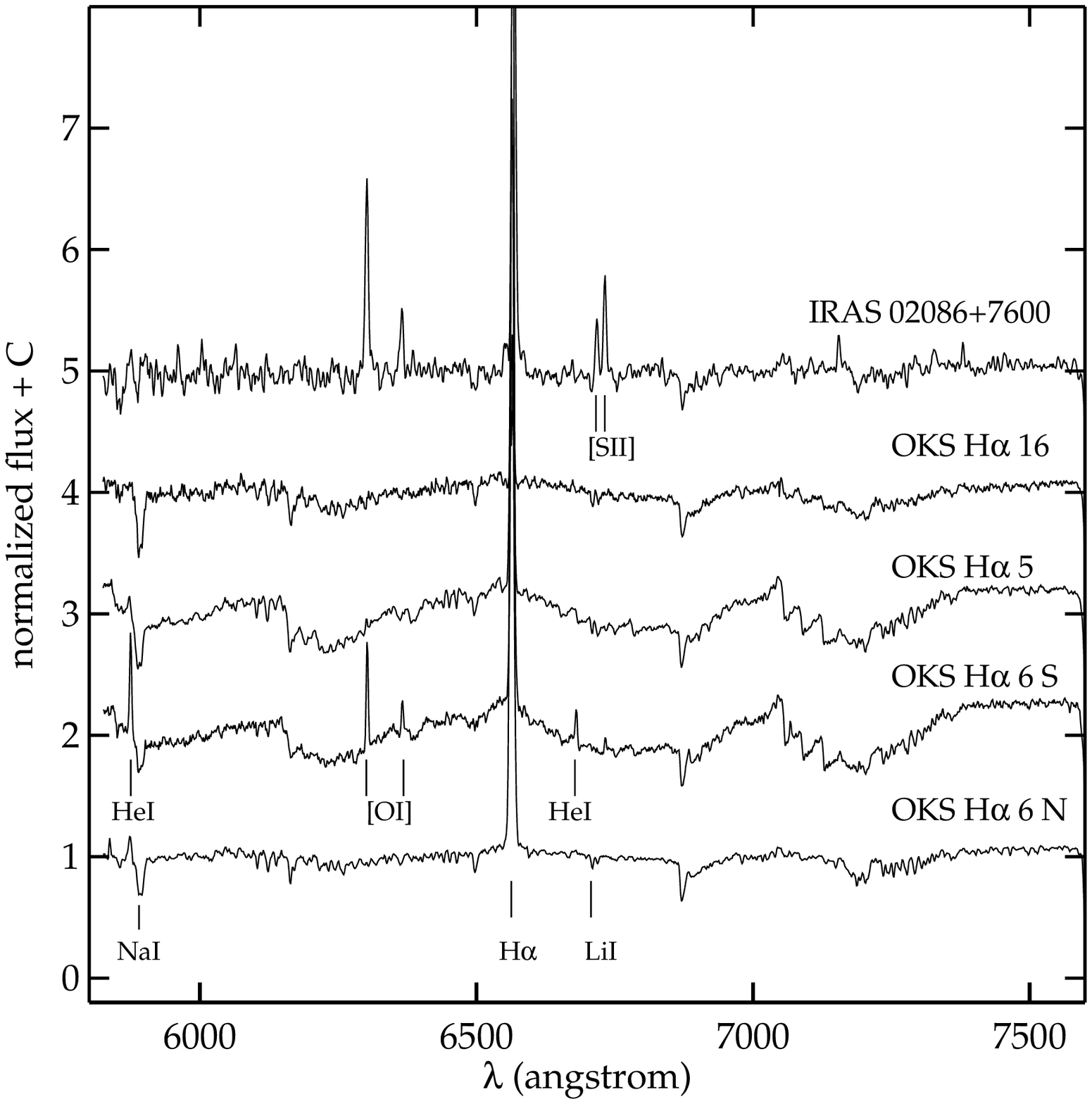}\includegraphics[width=8cm]{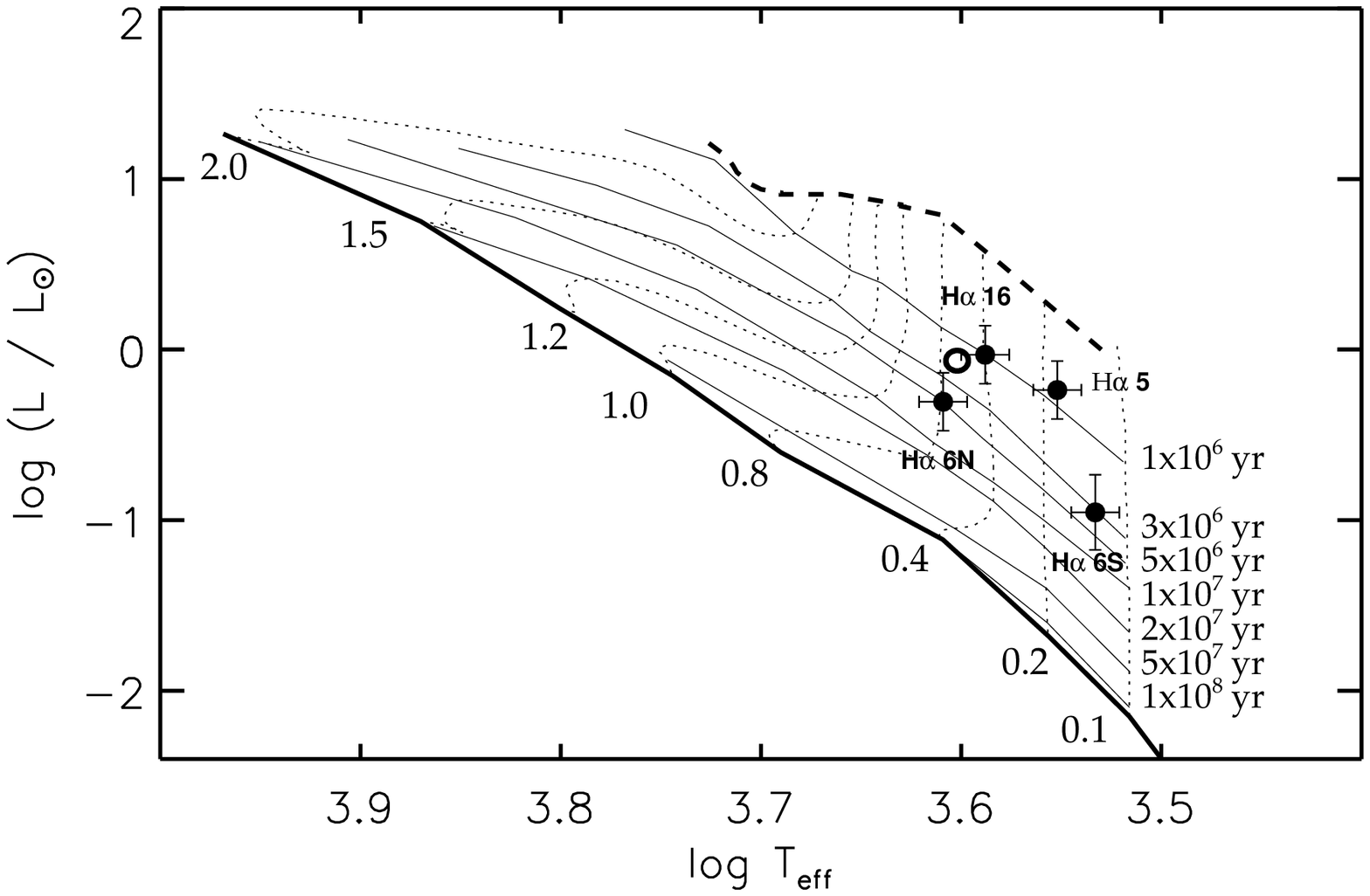}}
\caption{{\it Left\/}: Optical spectra of the young stellar objects in Lynds 1333. 
{\it Right\/}: Location of the same stars in the HRD, together with the pre-main-sequence evolutionary tracks and isochrones published by Palla \& Stahler (1999).}
\label{fig_1333_2}
\end{figure*}

The positions of the young stars mark two star-forming clumps along the filamentary cloud complex. Compared to other nearby star forming regions these star forming clumps are very small, similar to those found at high galactic latitudes (24\,M$_\odot$ and 15\,M$_\odot$, respectively; Obayashi et~al., 1998). The filamentary morphology of the cloud complex resembles the  L\,1495--B\,211--B\,213--B\,216 system in Taurus, but the separation of dense clumps along the filament is larger, and the star formation efficiency is smaller in L\,1333 than in Taurus. {\it Fig.~\ref{fig_map1333}} shows the distribution of the visual extinction in the region containing  L\,1333 (Dobashi et~al., 2005). The position of the dark cloud and the young 
stellar objects are marked. {\it Fig.~\ref{fig_1333_2}} shows the optical spectra of the young stars, obtained with the Nordic Optical Telescope, as well as their positions in the $\log T_\mathrm{eff}$  vs. $\log L$ diagram. 

Like in IC\,2118, the star forming cores of L\,1333 are significantly smaller than those of other well known nearby star forming regions (e.g. Taurus, Ophiuchus, Chamaeleon, Lupus). Interestingly, the young stars are located on the high galactic latitude side of the cloud, so that the oldest member of the group lies farthest from the cloud. Star formation in such small clumps is thought to be assisted by an external trigger. The observed morphology of the L\,1333 region suggests that the source of the trigger lies probably at the higher galactic latitudes (Kun et~al., 2006).

\CoKonSubsection{The Cepheus flare giant molecular complex}
\label{sec4.1.4}
 
The interstellar matter in the Cepheus at $100^{\rm o} < l < 120^{\rm o}$ and
$b > 10^{\rm o}$ is distributed over large spatial and velocity range (see e.g. Heiles, 1967; Lebrun, 1986; Yonekura et~al., 1999). Molecular gas has been observed in the radial velocity interval $-$15\,km\,s$^{-1} < v_{LSR} <$ +5\,km\,s$^{-1}$ (Grenier et~al., 1989; Yonekura et~al., 1999).
The nearby giant molecular complex of the Cepheus flare is comparable
in mass with the Taurus, Ophiuchus, and Chamaeleon complexes.

I determined the distances of the absorbing clouds using optical
star counts (Kun, 1998), and found three absorbing layers,
located at 200, 300 and 450 pc from the Sun, and equally
at $z \approx $ 90\,pc from the galactic plane. Farther away, at 
600--800\,pc from us, the outer regions of the associations Cep~OB2 
and Cep~OB3 are projected on the southern part of the Cepheus flare.
Our study of A-type stars of the region having far-infrared excess
confirmed this space distribution of the clouds (Kun, Vink\'o, \& Szabados, 2000). This morphology  suggests that the clouds are part of a larger
interstellar structure parallel to the galactic plane. Recently, Olano, Meschin, \& Niemela (2006) have shown that the interstellar matter in the
Cepheus flare is distributed over the surface of an expanding shell.
The presence of the shell can account for the different distances of
the dark clouds of the region.

I detected more than a hundred candidate pre-main-sequence stars over an
area of some 200 square degrees, covered by the clouds. Their distribution suggests that the three cloud complexes differ from each other in star forming activity. No high-mass star formation can be observed in the Cepheus flare region. Low-mass YSOs can be found in the component at 200\,pc (L\,1228) and in the 300\,pc component (NGC\,7023, L\,1235, L\,1251). The most distant component of the region can be found at the southern boundary of the complex. We searched for intermediate-mass pre-main sequence stars among the A and B type stars of the region exhibiting infrared excess, and identified a new Herbig~Ae star, BD\,$+68^{\mathrm o}1118$ (Kun, Vink\'o, \& Szabados, 2000). This star, together with a neighbouring HAe star HD\,203024, is projected on a relatively transparent region of the
cloud complex close to the star forming globule L\,1177 (CB~230).

Spectroscopic follow-up observations of the candidate pre-main sequence stars
were carried out using the 2.2\,m telescope of Calar Alto Observatory. $BVR_CI_C$ photometric observations of the same stars are in progress. The first results are published by Eredics \& Kun (2003).

\CoKonSubsection{L\,1340: A region of intermediate-mass star formation}
\label{sec4.3}

It is well known that high mass stars are born as members of dense clusters in giant molecular clouds, whereas small, cold  cores give birth to one or a few solar type stars. The transition from isolated to clustered mode of star formation occurs smoothly in  molecular clouds forming intermediate mass stars.

Lynds~1340 is a  molecular cloud in Cassiopeia, near (l,b)=(130,11), and associated with the reflection nebula DG~9 (Dorschner \& G\"urtler, 1966) illuminated by B and A-type stars. The small nebulosities RNO~7, 8, and 9 (Cohen, 1980), associated with the cloud, are probably signposts of recent star formation. We studied the structure and young stellar objects of Lynds~1340 in order to find how this birthplace of Herbig Ae/Be stars fits into the sequence of star forming environments. 

\begin{figure*}[htb]
\centerline{\includegraphics[width=14cm]{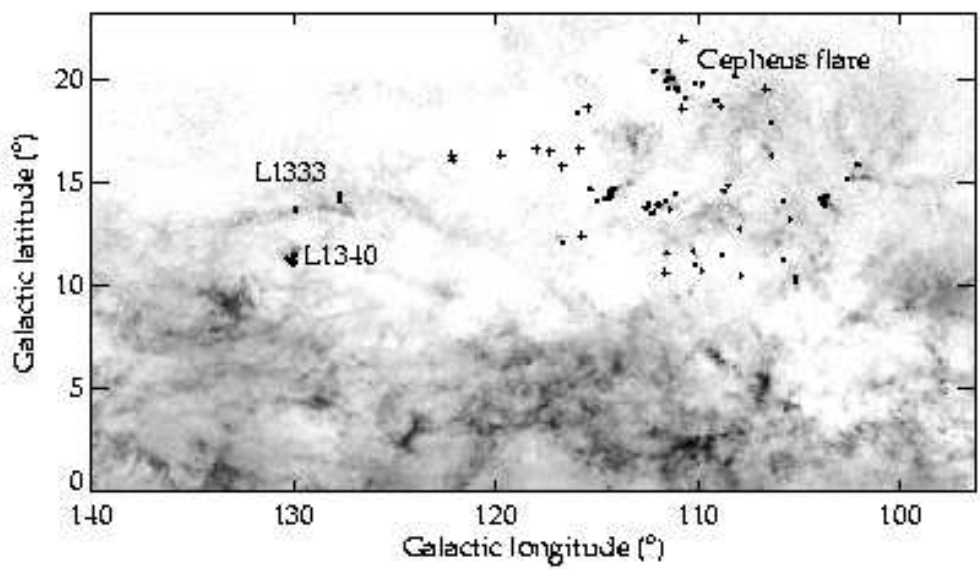}}
\vskip -7mm
\caption{Distribution of the visual extinction and the young stars in the Cepheus--Cassiopeia region.}
\label{fig_map}
\end{figure*}

\vfill\eject
The basic properties of the cloud are studied  within the framework of the 
collaboration between the Konkoly Observatory and the Radio Astronomy Laboratory of Nagoya University (Kun et~al., 1994). $^{13}$CO and C$^{18}$O maps of the cloud, obtained with the 4-m radio telescope of Nagoya University, distance determination, and a list of the candidate young stellar objects have been presented. The distribution of $^{13}$CO has revealed  three clumps within the cloud, each associated with a number of  {\it IRAS\/} point sources and H$\alpha$ emission stars. The total mass of the molecular cloud is some 1300 $M_\odot$. Follow-up spectroscopic, as well as optical and infrared photometric observations of L\,1340 revealed that a star as massive as some 6\,$M_\odot$ has been born in the cloud. This Herbig~Be star is a member of the embedded cluster RNO~7, consisting of some 25 stars. Follow-up radio observations in the 1.3\,cm inversion lines of the ammonia molecule have shown that several dense cores, future sites of star formation are embedded in L\,1340 (Kun et~al., 2003). Our results inspired several further studies of this interesting star forming region. For instance, Nanda Kumar et~al. (2002) and Magakian et~al. (2003) discovered several Herbig--Haro objects powered by the young stars embedded in the cloud. Submillimeter observations by O'Linger et~al. (2006) revealed protostellar objects
at very early stages of star formation.

The surface distribution of the young stellar objects in the Cepheus--Cassiopeia region at latitudes  $b > +10^\mathrm{o}$, discovered with the Schmidt telescope, is displayed in {\it Fig.~\ref{fig_map}} on the large-scale extinction map of the region (Dobashi et~al., 1995).

\CoKonSection{Future prospects}

The newly discovered nearby star forming regions are good targets for
more sensitive observations, aimed at revealing the whole stellar populations
born in them. Their comparison with the few well-known nearby star
forming regions (Orion, Taurus, Chamaeleon, Lupus) may shed light on hidden
laws of star forming processes. Further spectroscopic and photometric studies 
of the pre-main-sequence stars discovered will allow us to 
study some environmental effects of early stellar evolution.
\medskip

\CoKonAck The researches presented in this paper have been supported by the OTKA grants T\,022946, T\,034584, and T\,049082.

\CoKonReferences

Allen C. W., 1973, {\it Astrophysical Quantities}, Athlone Press, London

Brown A.G.A., Hartmann D., \& Burton W., 1995, {\it A\&A}, {\bf 300}, 903

Cohen M., 1980, {\it AJ}, {\bf 85}, 29

Cox A. N. (editor), 1999, {\it Allen's Astrophysical Quantities, Fourth Edition\/}, Springer

Cutri R. M., Skrutskie M. F., van Dyk S., et al., 2003, VizieR On-line Data Catalog: II/246

Dobashi K., Uehara H., Kandori R., Sakurai T., Kaiden M., Umemoto T., \& Sato F., 2005, {\it PASJ}, {\bf 57}, S1

Dorschner J. \& G\"urtler J., 1966, {\it AN}, {\bf 289}, 65

Dunst L., 1929, {\it Mitt Sternw. ung. Akad. Wiss.}, Budapest, No.\,1

Eredics M. \& Kun M., 2003, in {\it The interaction of stars with their
environments II\/}, eds. Cs. Kiss et al., Comm. Konkoly Obs., Budapest,
No.\,103, p. 27

Franco G. A. P., 1989, {\it A\&A}, {\bf 223}, 313

Grenier I. A., Lebrun F., Arnaud M., et al., 1989, {\it ApJ}, {\bf 347}, 231

Hartmann D. \& Burton W. B., 1997, {\it Atlas of Galactic Neutral Hydrogen\/}, Cambridge Univ. Press

Heiles C., 1967, {\it ApJS}, {\bf 15}, 97

Kawamura A., Kun M., Onishi T., Vavrek R., Domsa I., Mizuno A., \& Fukui Y., 2001, {\it PASJ}, {\bf 53}, 1097

Kirkpatrick J. D., Henry T. J., \& McCarthy D. W., 1991, {\it ApJS}, {\bf 77}, 417

Klessen R. S., Heitsch F., \& Mac Low M.-M., 2000, {\it ApJ}, {\bf 535}, 887

Kun M., 1998, {\it ApJS}, {\bf 115}, 59

Kun M. \& Prusti T., 1993, {\it A\&A}, {\bf 272}, 235

Kun M., Obayashi A., Sato F., Yonekura Y., Fukui Y., Bal\'azs L. G., \'Abrah\'am P., Szabados L., \& Kelemen J., 1994, {\it A\&A}, {\bf 292}, 249

Kun M., Vink\'o J., \& Szabados L., 2000, {\it MNRAS}, {\bf 319}, 777

Kun M., Aoyama H., Yoshikawa N., Kawamura A., Yonekura Y., Onishi T., \& Fukui Y., 2001, {\it PASJ}, {\bf 53}, 1063

Kun M., Wouterloot J. G. A., \& T\'oth L. V., 2003, {\it A\&A}, {\bf 398}, 169

Kun M., Prusti T., Nikoli\'c S., Johansson L. E. B., \& Walton N. A., 2004, {\it A\&A}, {\bf 418}, 89

Kun M., Nikoli\'c S., Johansson L. E. B., Balog, Z., \& G\'asp\'ar A., 2006, {\it MNRAS}, {\bf 371}, 732

Laureijs R. J., Fukui Y., Helou G., Mizuno A., Imaoka K., \& Clark F. O., 1995, {\it ApJS}, {\bf 101}, 87

Lebrun F., 1986, {\it ApJ}, {\bf 306}, 16

Lehtinen K., Mattila K., Lemke D., Juvela M., Prusti T., \& Laureijs R., 2003, {\it A\&A}, {\bf 398}, 571

Magakian T. Yu., Movsessian T. A., \& Nikogossian E. H., 2003, {\it Astrophysics}, {\bf 46}, 1

Magnani L., Blitz L., \& Mundy L., 1985, {\it ApJ}, {\bf 295}, 402

Mart\'{\i}n E. L. \& Kun M., 1996, {\it A\&AS}, {\bf 116}, 467 

Nanda Kumar, M. S., Anandarao B. G., \& Yu K. C., 2002, {\it AJ}, {\bf 123}, 2583 

Obayashi A., Kun M., Sato F., Yonekura Y., \& Fukui Y., 1998, {\it AJ}, {\bf 115}, 274

Olano C.A., Meschin P.I., \& Niemela V. S., 2006, {\it MNRAS}, {\bf 369}, 867

O'Linger J., Moriarty-Schieven G. H., \& Wolf-Chase G. A., 2006, submitted

Palla F. \& Stahler S. W., 1999, {\it ApJ}, {\bf 525}, 772

Preibisch T., Guenther E., \& Zinnecker H., 2001, {\it AJ}, {\bf 121}, 1040

T\'oth L. V., Hotzel S., Krause O., Lemke D., Kiss Cs., \& Mo\'or A., 2003, 
in {\it The interaction of stars with their environments II\/}, eds. Cs. Kiss et al., Comm. Konkoly Obs., Budapest, No. 103, p. 45

Yonekura Y., Dobashi K., Mizuno A., et al., 1997, {\it ApJS}, {\bf 110}, 21

\CoKonEndreferences

\CoKonChapEnd

\end{document}